\documentclass[10pt, conference]{IEEEtran}
\usepackage[utf8]{inputenc}
\usepackage{graphicx}
\usepackage{amsmath}
\usepackage{amssymb}
\usepackage{booktabs}
\usepackage{hyperref}
\usepackage{enumitem}
\usepackage{caption}
\usepackage{colortbl}
\usepackage{xcolor}
\usepackage{array}
\usepackage{float}

\title{Enhancing Law-Enforcement Audio Transcription: A LoRA-Based Adaptation of Whisper for BWC Footage}

\author{
\IEEEauthorblockN{
    \begin{tabular}{ccc}
        Vivek Senthil & Ernest Fokoué \\
        \texttt{vs9589@g.rit.edu} &  \texttt{epfeqa@rit.edu}
    \end{tabular}
}
\IEEEauthorblockA{
    Rochester Institute of Technology, Rochester, NY, USA
}
}

\begin{document}

\maketitle

\begin{abstract}
Body-worn camera (BWC) footage is a cornerstone of modern law enforcement evidence, yet its utility is often bottlenecked by the unstructured nature of audio data. Standard Automatic Speech Recognition (ASR) models frequently fail in this domain due to extreme environmental noise and specialized law-enforcement lexicons. This paper presents a parameter-efficient fine-tuning approach using Low-Rank Adaptation (LoRA) on the OpenAI Whisper-base model. By training only 0.3\% of the model's total parameters, we achieved a 39.7\% relative reduction in Word Error Rate (WER), outperforming both zero-shot and fully fine-tuned baselines. Furthermore, an ablation study on LoRA ranks reveals that lower-rank adaptations ($r=8$) are optimal for capturing domain-specific acoustic patterns without overfitting to the noisy distributions inherent in BWC recordings.
\end{abstract}

\section{Introduction}
Law enforcement agencies across the United States, such as the Rochester Police Department (RPD), currently manage and store petabytes of Body Worn Camera (BWC) footage. While these recordings provide critical transparency and accountability, the sheer volume of data makes comprehensive manual review virtually impossible. At present, this vast archive remains fundamentally "unstructured" data. Locating a specific ten-second interaction for evidence, judicial review, or internal monitoring can require hours of exhaustive manual labor. 

This project addresses a critical bottleneck: the urgent need for automated, high-accuracy transcription to make the justice system more efficient. By enabling accurate and scalable transcription, departments can analyze police-public interactions, monitoring for respectfulness, de-escalation tactics, or implicit bias, without the prohibitive cost of human review. The ultimate goal of this research is to investigate whether domain-specific fine-tuning can transform foundation models like OpenAI's Whisper \cite{electronics13214227} into a robust tool for this highly challenging acoustic environment.

\section{Domain Challenges and Motivation}
The primary motivation for this work is the sheer inadequacy of manual transcription in the face of exponentially growing data volumes. However, automating this process exposes a critical linguistic barrier that standard ASR systems are fundamentally unequipped to handle.

\subsection{The Out-of-Vocabulary (OOV) Barrier}
A central challenge in transcribing law‑enforcement audio is the prevalence of Out‑of‑Vocabulary (OOV) terms. Police communication is saturated with tactical codes, legal jargon, unit identifiers, and region‑specific slang that rarely appear in the large‑scale, web‑scraped corpora used to train foundation models like Whisper. When confronted with this specialized lexicon, generalized ASR systems attempt to force unfamiliar acoustic patterns into the closest match within their existing vocabulary. This results in severe semantic drift and misrecognizing “10‑52,” “Mirandize,” or “Signal 13” as unrelated everyday words, and ultimately erodes the operational meaning of the transcript. The inability to correctly model these domain‑specific terms represents a structural limitation of zero‑shot ASR and motivates the need for targeted adaptation.

\section{Methodology and Formulation}
This research operates at the intersection of Automatic Speech Recognition (ASR) and Natural Language Processing (NLP). 

\subsection{Sequence-to-Sequence Mapping Framework}
We define the transcription problem as a sequence-to-sequence mapping task. We employ a Transformer-based Encoder-Decoder architecture, specifically utilizing the OpenAI Whisper model (Fig. 1). The model consumes raw audio, which is converted into log-Mel spectrograms, and generates corresponding text tokens. 

\begin{figure}[htbp]
    \centering
    \includegraphics[width=0.9\linewidth]{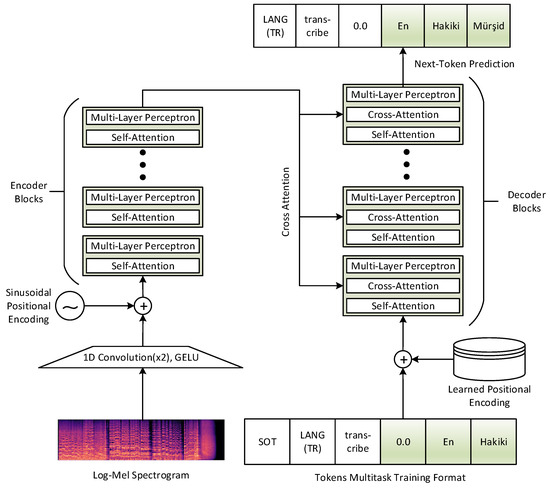}
    \caption{Transformer-based Encoder-Decoder architecture of the Whisper model \cite{electronics13214227}.}
    \label{fig:whisper_arch}
\end{figure}

Our methodology follows a generative, parametric approach through supervised fine-tuning. Under this framework, the model is trained to minimize the loss function by predicting the next token in a sequence, explicitly conditioned on the acoustic features of the law enforcement audio. 

\subsection{Parameter-Efficient Fine-Tuning via LoRA}
Given the model's massive scale, full retraining is computationally expensive, requires vast amounts of memory, and risks "catastrophic forgetting." Therefore, we implement Parameter-Efficient Fine-Tuning (PEFT) via LoRA \cite{hu2021lora}. 

Low-Rank Adaptation introduces an efficient parameterization for fine-tuning large neural networks by decomposing weight updates into a pair of low-rank matrices. Instead of updating the full pretrained weight matrix $W \in \mathbb{R}^{d \times k}$, LoRA constrains the update to a rank-$r$ subspace, where $r \ll \min(d, k)$. 

\begin{figure}[htbp]
    \centering
    \includegraphics[width=0.8\linewidth]{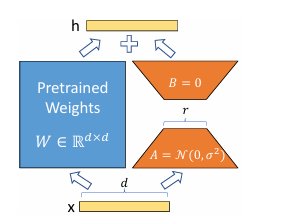}
    \caption{Low-Rank Adaptation framework injecting matrices A and B into frozen weights \cite{hu2021lora}.}
    \label{fig:lora}
\end{figure}

The trainable update is mathematically expressed as:
\begin{equation}
\Delta W = BA
\end{equation}
where $A \in \mathbb{R}^{r \times k}$ and $B \in \mathbb{R}^{d \times r}$ are the low-rank factors. During fine-tuning, the original pretrained weight $W$ remains entirely frozen, and only matrices $A$ and $B$ are optimized through backpropagation. 

The forward pass becomes:
\begin{equation}
h = Wx + \Delta Wx = Wx + BAx
\end{equation}
To ensure that LoRA does not alter the pretrained model's behavior at initialization, the weights are specifically initialized with $B=0$ and $A$ sampled from a random Gaussian distribution $\mathcal{N}(0, \sigma^2)$. Therefore, at the start of training, $\Delta W = BA = 0$, maintaining the original zero-shot performance. For this work, we restrict LoRA updates to the query and value projection layers (\texttt{['q\_proj', 'v\_proj']}).

\section{Experimental Setup and Dataset Curation}
\subsection{Data Acquisition and Filtering}
The data collection phase aimed to isolate authentic, noisy audio environments that mimic or are actual body-cam footage. 
\begin{itemize}[leftmargin=*]
    \item \textbf{Data Collection:} A curated set of 294 videos with existing human‑generated transcripts was assembled from publicly available broadcast content.
    \item \textbf{Filtering Pipeline:} Non-relevant content such as courtrooms and controlled interrogations was manually filtered out to ensure evaluation exclusively on challenging BWC audio.
    \item \textbf{Final Corpus:} We retained a highly refined corpus of 53 body-worn camera videos.
\end{itemize}

\begin{figure}[htbp]
    \centering
    \includegraphics[width=0.9\linewidth]{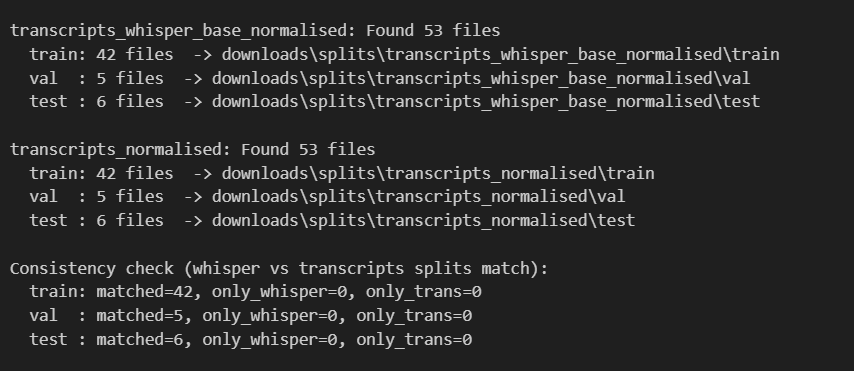}
    \caption{Consistency check confirming exact split matches between audio and transcripts for the 53 BWC files.}
    \label{fig:data_split}
\end{figure}

As shown in Fig. \ref{fig:data_split}, the dataset was strictly partitioned to prevent data leakage, resulting in 42 Train, 5 Validation, and 6 Test files (an 8:1:1 ratio) \cite{senthil2026asrwhisperfinetuning}.

\subsection{Preprocessing and Normalization Pipeline}
Ground truth data required extensive normalization. Human-generated ground truth data was normalized to remove non-alphanumeric characters while preserving the VTT (Video Text Tracks) timestamps, allowing us to accurately chunk the continuous audio into distinct, trainable segments.

\subsection{Training Hardware and Workflow}
\begin{figure}[htbp]
    \centering
    \includegraphics[width=\linewidth]{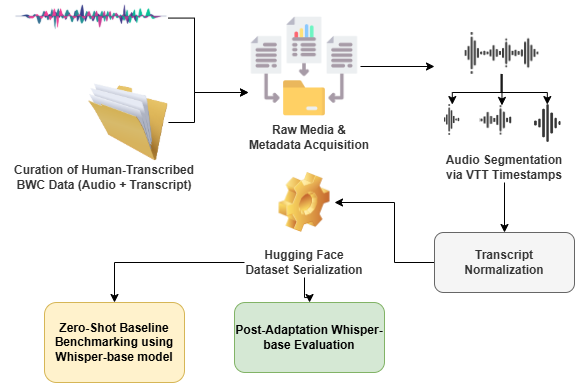}
    \caption{Whisper Adaptation and Benchmarking Workflow.}
    \label{fig:workflow}
\end{figure}

The adaptation workflow (Fig. \ref{fig:workflow}) manages the transition from raw media acquisition to final evaluation. 

The model was fine‑tuned on RIT’s Research Computing infrastructure \cite{https://doi.org/10.34788/0s3g-qd15}, leveraging NVIDIA A100 20GB GPUs and the optimized ASR training scripts described in \cite{theodb2023}. We employed LoRA with varying ranks ($r=8, 16, 32$), a scaling factor ($\alpha$) of 32, and a dropout rate of 0.05 to mitigate overfitting. 

\subsection{Evaluation Metrics}
Success was strictly quantified using the Word Error Rate (WER) metric:
\begin{equation}
WER = \frac{S + D + I}{N}
\end{equation}
where $S$ represents substitutions, $D$ deletions, $I$ insertions, and $N$ the total number of words.

\section{Results and Baselines Analysis}
\subsection{Baseline Performance Assessment}
We benchmarked the LoRA-adapted model against a Zero-shot Whisper-base model and a Fully Fine-Tuned Whisper-base model. 

\begin{table}[h]
\centering
\caption{Data Split}
\begin{tabular}{l c}
\toprule
\textbf{Split} & \textbf{Files} \\
\midrule
Train & 42  \\
Val & 5  \\
Test & 6  \\ \bottomrule
\end{tabular}
\end{table}

\subsection{LoRA Performance and Rank Ablation}
The integration of LoRA successfully bridged the domain gap. As detailed in Table II, our proposed LoRA configurations drastically outperformed the baselines.

\begin{table}[h]
\centering
\caption{Comparative Analysis of LoRA Ranks vs. Baselines}
\begin{tabular}{l r c c} 
\toprule
\textbf{Model Configuration} & \textbf{Trainable Params} & \textbf{Rank ($r$)} & \textbf{Avg. WER} $\downarrow$ \\ 
\midrule
Whisper-base (Zero-shot) & 0 & - & 0.6194 \\ 
Full Fine-Tuned & 99,148,800 & - & 0.5874 \\ \midrule
\textbf{LoRA Optimized} & \textbf{294,912} & \textbf{8} & \textbf{0.3733} \\ 
LoRA Optimized & 589,824 & 16 & 0.3793 \\ 
LoRA Optimized & 1,179,648 & 32 & 0.3848 \\ 
\bottomrule 
\end{tabular}
\end{table}

The optimal configuration was achieved with a LoRA rank of $r=8$, representing a massive 39.7\% relative WER reduction compared to the out-of-the-box base model. 

\section{Scenario and Qualitative Analysis}
\subsection{Scenario Sensitivity}
\begin{figure}[htbp]
    \centering
    \includegraphics[width=\linewidth]{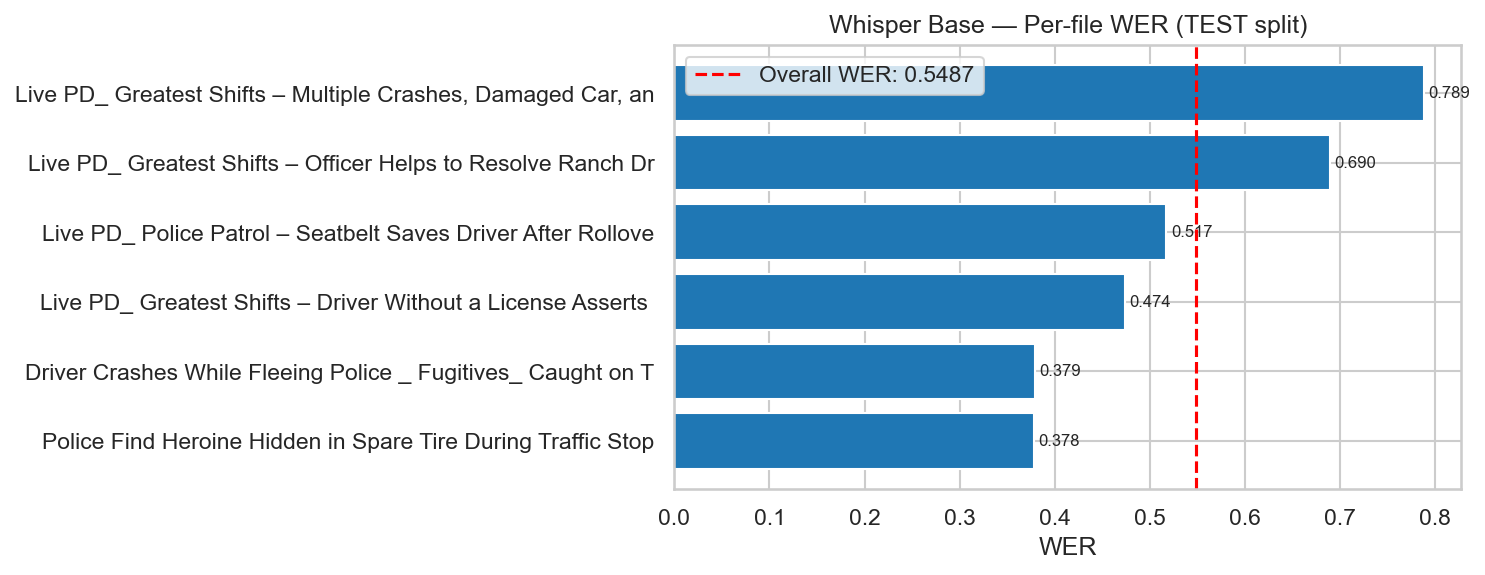}
    \caption{WER variance across test scenarios.}
    \label{fig:wer_test}
\end{figure}

In the test set, the model proved highly sensitive to environmental factors. As shown in Fig. \ref{fig:wer_test}, it performed best on routine, simple traffic stops (0.378 WER) but failed on complex crash scenes (0.789 WER). 

\subsection{Qualitative Examples and OOV Resolution}
A review of qualitative outputs demonstrates LoRA's effectiveness in resolving the Out-of-Vocabulary problem. 

\begin{table}[h]
\centering
\caption{Selected Prediction Comparisons illustrating OOV Handling}
\resizebox{\columnwidth}{!}{%
\begin{tabular}{>{\raggedright\arraybackslash}p{0.3\linewidth} >{\raggedright\arraybackslash}p{0.3\linewidth} >{\raggedright\arraybackslash}p{0.3\linewidth}}
\toprule
\textbf{Reference Ground Truth} & \textbf{Baseline (Zero-shot)} & \textbf{LoRA Prediction (r=8)} \\
\midrule
on North Ammon Road headed south & on Am south & on North Ammon Road headed south \\
\midrule
They just wrecked at the roundabout & They just the the & They just wrecked at the roundabout \\
\midrule
Can I get county units out here Expedite & Can Can I.. & Can I get county units or here expedite \\
\bottomrule
\end{tabular}
}
\end{table}

In the baseline outputs, the model heavily deletes information, turning "North Ammon Road" into "Am south" and failing to interpret dispatch requests like "Expedite". The $r=8$ LoRA model successfully learned these tactical phrases.

\section{Discussion}
Several key technical insights arise from these experimental results:

\begin{itemize}[leftmargin=*]
    \item \textbf{Optimal Rank Efficiency and Diminishing Returns:} Our ablation study revealed that increasing the LoRA rank beyond 8 ($r=16, r=32$) led to slight performance degradation (rising from 0.3733 to 0.3848 WER). Lower-rank updates are sufficient for capturing BWC acoustic patterns; higher ranks may cause overfitting to the severe noise artifacts.
    \item \textbf{Parameter Economy:} We achieved a 39.7\% improvement in transcription accuracy while updating only 0.3\% of the total parameters (294,912 compared to over 99 million for full fine-tuning).
\end{itemize}

\section{Conclusion and Future Work}
This capstone research successfully demonstrated that foundation ASR models can be effectively adapted for the harsh acoustic environments of law enforcement using Low-Rank Adaptation. By employing an $r=8$ LoRA configuration targeting the attention projection matrices, we achieved a near 40\% reduction in Word Error Rate, bridging the domain gap while maximizing computational parameter economy. Future work must investigate why higher adaptation ranks introduced noise into the transcription process and focus on implementing hybrid audio preprocessing pipelines.

\section*{Acknowledgment}
Supported by a U.S. Department of Justice grant (15PBJA-22-GG-03328-BWCx), awarded to the City of Rochester.

\bibliographystyle{IEEEtran}
\bibliography{references}

\end{document}